\documentstyle[aps,epsf,prd,multicol]{revtex}

\begin{document}
{\baselineskip-4pt
\font\yitp=cmmib10 scaled\magstep2
\font\elevenmib=cmmib10 scaled\magstep1  \skewchar\elevenmib='177
\leftline{\baselineskip20pt
\hspace{8mm} 
\hspace{10mm} 
\vbox to0pt
   { {\yitp\hbox{Osaka \hspace{1.5mm} University} }
     {\large\sl\hbox{{Theoretical Astrophysics}} }\vss}}

\rightline{\large\baselineskip20pt\rm\vbox to20pt{
\baselineskip14pt
\hbox{OU-TAP-191}
\hbox{YITP-03-2}
\vspace{1mm}
\hbox{\today}\vss}}%
}
\vskip8mm
\begin{center}{\large \bf Generation of dark radiation in the bulk
inflaton model} 
\end{center}
\vspace*{4mm}
\centerline{\large 
Takahiro Tanaka$^1$ and Yoshiaki Himemoto$^{1 ,\, 2}$}
\vspace*{4mm}
\centerline{\em $^1$Yukawa Institute for Theoretical Physics,
Kyoto University, Kyoto 606-8502, Japan}
\centerline{\em $^2$Department of Earth and Space Science, Graduate 
School of Science}
\centerline{\em Osaka University, Toyonaka 560-0043, Japan} 

\begin{abstract}
We investigate the dynamics of a bulk scalar field with various decay 
channels in the Randall-Sundrum infinite braneworld scenario. 
A bulk scalar field in this scenario has a quasilocalized mode 
which dominates the late-time behavior near the brane.
As for this mode, an interesting point is the presence of 
dissipation caused by the escape of the energy in the direction 
away from the brane, even if the bulk scalar field does not have 
the interaction with the other bulk fields in the bulk and fields on the
brane. We can interpret that this lost energy is transferred to 
the dark radiation.  
We show that such an effective 4-dimensional description for a bulk 
scalar field is valid including the various 
processes of energy dissipation. 



\end{abstract}
\vskip5mm
\begin{multicols}{2}

\section{Introduction}
The braneworld scenario has provided a new scheme to derive
4-dimensional effective 
theory from a fundamental higher 
dimensional one~\cite{Horava:1996qa}. 
In particular, the second model proposed by Randall and Sundrum 
(RS2) has demonstrated 
an exciting possibility that the extra dimensions may 
extend infinitely \cite{Randall:1999vf}. 
A lot of work has been done on various aspects of this model. 
Extensive research based on relativity and cosmology, which is 
closely related to the issue we will discuss in this paper, has 
also been 
developed\cite{Shiromizu:2000wj,Garriga:2000bq,Garriga:2000yh,cosmology}.

The background spacetime of the original RS model is given by a 
$Z_2$-symmetric solution of the 5-dimensional Einstein 
equations with a negative cosmological constant. 
However, in the framework of 
string theory, it is more likely that there are various 
fields other than the graviton in the 5-dimensional bulk. 
Thus, 
the dynamics of a minimally coupled 5-dimensional
massive scalar field in the braneworld scenario of RS2 type 
has been studied, and it was shown that 
a bulk scalar field can play the role of 
inflaton field\cite{Himemoto:2001hu,Himemoto:2001nd,newronbun}.
We refer to such a model as the bulk inflaton model. 

Prior to the proposal of the bulk inflaton model,
Dubovsky, Rubakov, and Tinyakov
have analyzed the dynamics of a bulk scalar field with the mass $m$ 
on the RS2 Minkowski brane background~\cite{Dubovsky:2000am}.  
They have found that the bulk scalar field has a quasilocalized mode
which dominates the late-time behavior near the brane. 
When this mode dominates, the scalar field seen on the brane
oscillates with the frequency given by $m/\sqrt{2}$ 
and decays gradually. 
Besides this interesting intrinsic decay, 
when we try to construct a more realistic bulk inflaton model, 
we also need to take account of the dissipation due to the 
interaction with other bulk fields and fields on 
the brane{\footnote{The latter process was investigated in our previous
paper\cite{newronbun}.}}.
We discuss such decay processes in this paper,
focusing on the effect on the cosmic expansion 
law realized on the brane. 
Related discussions about the interaction between the bulk 
and the fields on the brane were given in \cite{interaction}. 
In particular, generation of dark radiation via collision of 
particles on the brane was discussed in \cite{DR}.  



This paper is organized as follows. 
In Sec.~II we will 
review the general formulas for the cosmic expansion law 
under the presence of bulk fields. 
In Sec.~III we will study various decay channels of the bulk scalar field, 
and show how the decay products affect the cosmic expansion 
law on the brane. 
Section IV is devoted to summary.

\section{cosmic expansion law in the generalized Randall-Sundrum 
braneworld}

We consider the braneworld scenario of the RS2 type,
starting with the 5-dimensional Einstein equations
\begin{equation}
  R_{ab}-{1\over2}g_{ab}R+\Lambda_5g_{ab}
=\kappa_5^2\left[{T}_{ab}+S_{a b} \delta(r-r_0)\right]\,,
\label{bulkeq}
\end{equation}
where $r$ is the proper coordinate normal to the brane, and
the brane is located at $r=r_0$ at the 
fixed point of the $Z_2$ symmetry.
$S_{ab}$ represents the energy-momentum tensor composed of 
the brane tension $\sigma$ and 
the contribution of matter fields 
confined on the brane $\tau_{ab}$:
\begin{equation}
S_{a b} = -\sigma q_{a b}+\tau_{ab}\,,
\label{matter}
\end{equation}
where $q_{ab}$ is the metric induced on the brane.
We choose 
$\sigma:=\sqrt{-6\Lambda_5}/\kappa_5^{2}=6/\kappa_5^{2}\ell$ 
so that the Randall-Sundrum flat brane is realized 
as the ground state. Here $\ell$
is the bulk curvature radius.
$T_{ab}$ is the effective 5-dimensional energy-momentum tensor 
of the bulk fields. 
Here we do not specify the form of $T_{ab}$.

It is now well known that the 4-dimensional 
effective Einstein equation is derived from Eq.(\ref{bulkeq}) 
as\cite{Shiromizu:2000wj} 
\begin{equation}
  {}^{(4)}G_{\mu\nu}=
\kappa_4^2(\tau_{\mu\nu}
           +\tau_{\mu\nu}^{(\pi)}+\tau^{(B)}_{\mu\nu}+\tau_{\mu\nu}^{(E)}
            ),  
\label{4dEinstein}
\end{equation}
where $\kappa_{4}^{2} = {\kappa_5^2/\ell}$ 
and 
\begin{eqnarray}
\tau_{\mu \nu}^{(\pi)} &=& 
 -{\kappa_5^2 \ell\over 24}\bigl(6\tau_{\mu \alpha}\tau_{\nu}^{\ \alpha}
 -2\tau \tau_{\mu \nu} 
 -q_{\mu \nu}(3\tau_{\alpha \beta}\tau^{\alpha\beta}
  -\tau^{2})\bigr),\hspace{-1cm}
\nonumber\\
  \tau_{\mu\nu}^{(B)} &= &{2\ell\over 3}
   \left[{T}_{ab} q^a{}_\mu q^b{}_\nu
          +q_{\mu\nu}\left({T}_{ab} n^a n^b
            -{1\over 4} {T}^{a}{}_a
             \right) \right]_{r=r_0},\hspace{-0cm}
\nonumber\\
\tau_{\mu \nu}^{(E)}&=& \,-{\ell \over \kappa_5^2} 
  \left[{}^{(5)}C_{a b c d}
\,n^{a}\,n^{c}\,q_{\mu}^{b}\,q_{\nu}^{d}\right]_{r=r_0}\,,
\label{eq4}
\end{eqnarray}
where $n^{a}$ is the unit vector pointing outward and  
normal to the brane. 
Here $\tau_{\mu \nu}^{(E)}$ is the energy-momentum tensor related to
the 5-dimensional Weyl tensor ${}^{(5)}C_{a b c d}$.



To discuss the braneworld cosmology, we assume the spatial homogeneity and
isotropy for $q_{\mu \nu}$ and 
for each component of the energy-momentum tensor.
We define the energy density of the 4-dimensional ordinary matter 
fields by $\rho:=u^\mu u^\nu \tau_{\mu\nu}$,
where $u^\mu$ is the timelike unit vector associated with comoving
observers on the brane. 
Similarly, we define 
$\rho^{(i)}:=u^\mu u^\nu \tau^{(i)}_{\mu\nu}$ 
for each component. 
Then, Eq.~(\ref{4dEinstein})
gives the Friedmann equation on the brane: 
\begin{eqnarray}
H^2 =  
{{\kappa _4^2} \over 3}\left[ 
\left(1+{\rho\over 2\sigma}\right)\rho+\rho^{(B)} 
 +\rho^{(E)}\right], 
\label{Hubble}
\end{eqnarray}
where $H$ is the Hubble parameter on the brane. 
Here we have used the fact $\rho^{(\pi)}= \rho^2/2\sigma$. 

Next we consider the equations for the evolution of the energy density.
From the 5-dimensional conservation law 
integrated over infinitesimally thin layer near the brane, 
the evolution equation for $\rho$ is obtained as 
\begin{equation}
  \dot \rho+4H\rho+H \tau^{\mu}_{~\mu}= 
  2 {T}_{ab} u^a n^b\vert_{r=r_0}\,, 
\label{cons4d}
\end{equation}
where an overdot means differentiation 
with respect to the cosmic proper time $t$. 

On the other hand, 4-dimensional Bianchi identity 
implies 
\begin{equation}
 \dot\rho^{({\rm{tot}})}+ 4H\rho^{({\rm{tot}})}
+H \tau^{({\rm{tot}})\mu}{}_{\mu}=0,
\end{equation}
where we have defined 
$\rho^{({\rm{tot}})}:=\rho+\rho^{(\pi)}+\rho^{(B)}+\rho^{(E)}$ 
and $\tau^{({\rm{tot}})}_{\mu\nu}$ in the same way. 
Then, using Eq.~(\ref{cons4d}) and the facts that
$\tau^{(\pi)\mu}{}_{\mu}= {(\rho/\sigma)}(\tau^{\mu}{}_{\mu}+2\rho)$, 
$ \tau^{(B)\mu}{}_{\mu} = 2 \ell\, {T}_{a b} n^a n^b$, and
$ \tau^{(E)\mu}{}_{\mu} = 0$, 
we find that 
the effective energy density of the bulk field 
$\rho_{\rm{eff}}:=\rho^{(B)}+\rho^{(E)}$ satisfies  
\begin{equation}
 \dot \rho_{\rm{eff}}+4H\rho_{\rm{eff}}
  =\left[-2\left(1+{\rho\over\sigma}\right){T}_{ab} u^a n^b  
  -2 H\ell \, {T}_{a b} n^a n^b\right]_{r=r_0}\,.
\label{cons5d1}
\end{equation}
The above Eqs.(\ref{Hubble}), (\ref{cons4d}), and (\ref{cons5d1})
completely determine the cosmic expansion law 
once 
we know the evolution of the energy-momentum tensor of the 
5-dimensional fields evaluated on the brane.

\section{dissipation of the bulk scalar field}

As a phenomenological model of a bulk scalar field with dissipation, 
we consider the field equation given by
\begin{equation}
 \Box_5\phi-m^{2}\phi
= 
\left(\Gamma_d \,\ell \, \delta \left(r-r_{0}\right)
+{\hat \Gamma_b(r) \over N(r)}\right)\dot{\phi}\,,
\label{kgv}
\end{equation}
where $m^{2}$ is the mass squared of a bulk scalar field and
$N(r)$ is the lapse function. 
Here $\Gamma_d$ represents the rate of the decay
on the brane, 
while $\hat\Gamma_b(r)$ is a function specifying 
the rate of energy release to the bulk. 
We call this $\phi$ field 
the bulk inflaton, although our main interest is 
not in the period of inflation but 
in the period of reheating after inflation. 
For a moment, we do not specify the functional form of $\hat\Gamma_b(r)$. 
In this setting,
we study the cosmic expansion law 
by analyzing the set of equations (\ref{Hubble}), (\ref{cons4d}), and
(\ref{cons5d1}).


\subsection{Dark radiation from escaping bulk inflaton}
First, we consider the simplest case in which 
there is no bulk field other than the bulk inflaton 
field $\phi$ that has no direct interaction to the fields on the
brane. 
Namely, $T_{ab}=T^{(\phi)}_{a b}$ and $\Gamma_d=\hat\Gamma_b=0$, 
where
\begin{equation}
T_{ab}^{(\phi)}
=\phi_{,a}\phi_{,b}-g_{ab}\left(
  {1\over 2}g^{cd}\phi_{,c}\phi_{,d}
  +{1\over 2} m^2\phi^2\right)
\label{EMtensor}
\end{equation}
is the energy-momentum tensor of the bulk inflaton field.  
Examining this simplest case is instructive to figure out 
an important difference between a bulk field and a field 
on the brane. 

According to the analyses of the dynamics of a massive bulk scalar field
on the RS2 flat brane background~\cite{Dubovsky:2000am} and 
on the de Sitter brane background~\cite{Himemoto:2001hu}, 
there is a quasilocalized mode which dominates the late-time behavior. 
This quasilocalized mode corresponds to a pair of the 
first poles appearing in the retarded Green function in the frequency domain. 
To summarize the results obtained in the above references, 
$\phi$ behaves as $\propto e^{[\pm i\omega-(3H/2)-(\Gamma/2)]t}$
with 
\begin{eqnarray}
  \omega \hspace{-5mm} &\approx & \hspace{-5mm} 
  \sqrt{m^2_{\mathrm {eff}}-{9\over 4}H^2}
\displaystyle\mathop
{\longrightarrow}_{H^2/m^2\to 0} m_{\mathrm {eff}}, 
\cr
  \Gamma \hspace{-5mm} &= & \hspace{-5mm} 
   \Gamma_{\rm{esc}} \approx  {m^2(m^2-4 H^2)\ell^4
          \over 8 \omega} {\rm Im}\left[\psi\left({1\over 2}
-{i\omega\over H}\right)\right]\cr
& ~~\displaystyle\mathop
{\longrightarrow}_{H^2/m^2\to 0}& m^3\ell^2\pi/8\sqrt{2},  
\label{basic}
\end{eqnarray}
under the conditions 
\begin{equation}
m \ell\,  \;{\rm{and}}\;H \ell \, \ll 1 .
\label{condition}
\end{equation}
Here $m_{\mathrm{eff}}^2=m^2/2$, and 
$\psi(z)\equiv \Gamma'(z)/\Gamma(z)$ is the poly-Gamma function. 
The bulk 
inflaton oscillates with the frequency 
$\omega$ and decays intrinsically even if we do not introduce 
dissipation effects due to the coupling to other fields. 
From the 5-dimensional point of view,
this decay of the amplitude of oscillation occurs as a 
result of the escape of the energy in the direction away from the brane. 
We note that the above expressions are valid only when 
$m_{\mathrm {eff}}^{2}/H^{2}> 9/4$. 
Here our discussion is therefore restricted to the reheating 
era after inflation.

Let us consider the effective energy density of the bulk inflaton field 
from the viewpoint of 4-dimensional observers.
In the present case, a $Z_2$-symmetric boundary condition implies 
$\phi_{,a}n^a=0$. Therefore we have
$ T^{(\phi)}_{ab} u^a n^b = 0$, and then 
the energy conservation equation~(\ref{cons4d}) reduces 
to the same equation as in the standard 4-dimensional 
cosmology:  
\begin{equation}
  \dot \rho+4H\rho+H \tau^{\mu}_{~\mu}= 0. 
\label{cons4d2}
\end{equation}
Furthermore, from Eq.~(\ref{cons5d1}), 
we find that 
$\rho_{\mathrm{eff}}=\rho^{(E)}+\rho^{(\phi)}$ satisfies  
\begin{eqnarray}
 \dot \rho_{\rm{eff}}+4H\rho_{\rm{eff}}
   =  -H\ell\left[
      \dot\phi^2-m^2\phi^2\right]_{r=r_0}. 
\label{cons5d}
\end{eqnarray}
This equation tells us that
the evolution of $\rho_{\rm{eff}}$ depends on the 
evolution of $\phi$ on the brane. 
Therefore, in order to determine the cosmic expansion law, 
we need to know the evolution of the bulk scalar field. 

Next, for comparison we propose a 4-dimensional model
composed of a scalar field $\Phi$ and the dark radiation in addition 
to the ordinary matter fields. 
We should stress that in the 4-dimensional 
model the dark radiation simply means a component of energy density 
decoupled from the ordinary matter fields.  
As for the ordinary matter fields, we use the same notations 
as in the case of the 5-dimensional model.
Then the conservation law for the ordinary matter fields is 
identical to 
Eq.~(\ref{cons4d2}).

We define a scalar field $\Phi$ which reproduces the late-time 
behavior of the value of $\phi$ on the brane besides an overall 
normalization factor. 
We assume that $\Phi$ has mass $m_{\mathrm{eff}}$ 
and decays only 
into the dark radiation with the decay width $\Gamma_{\rm{esc}}/2$. 
The equation of motion for the homogeneous part of the scalar field 
is given by 
\begin{eqnarray}
\ddot\Phi+(3H+\Gamma_{\rm{esc}})\dot\Phi+m_{\rm{eff}}^{2}\Phi
=0.
\label{infdynamics}
\end{eqnarray}
Using $\Gamma_{\rm{esc}}\ll m$, which follows 
from $m\ell\ll 1$, 
it is easy to see that $\Phi$ also behaves as 
$\propto e^{[\pm i\omega-(3H/2)-(\Gamma_{\rm{esc}}/2)]t}$ 
with the same $\omega$ in Eq.~(\ref{basic}). 

The Hubble equation is given by 
$
H^2 =  
({{\kappa _4^2} / 3}) ( 
\rho + \tilde\rho_{\mathrm{eff}}), 
$
with $\tilde\rho_{\mathrm{eff}}:=\rho^{(\Phi)} +\rho^{(D)}$,
where $\rho^{(\Phi)}$ and $\rho^{(D)}$
are the energy density of the 
$\Phi$ field and the dark radiation, respectively. 
Since the scalar field $\Phi$ and the dark radiation 
are decoupled from the ordinary matter fields by assumption, 
they separately satisfy the conservation law. 
The conservation law for $\Phi$ and the dark radiation 
reduces to
\begin{eqnarray}
\dot {\tilde \rho}_{\rm{eff}} +4H 
\tilde{\rho}_{\rm{eff}}
&=&-H\tau^{(\Phi)\mu}{}_{\mu} 
= -H\left(\dot\Phi^2-2m_{\mathrm{eff}}^2\Phi^2\right).  
\label{cons4dmodel}
\end{eqnarray}
Then, with simple identification 
\begin{equation}
 \Phi=\sqrt{\ell}\,\phi\,,
\label{Phidef}
\end{equation}
it is easy to see that 
the 4-dimensional and the 5-dimensional 
models presented above realize 
the same cosmic expansion in the law energy regime (\ref{condition}),  
in which the contribution from $\rho^{(\pi)}$ 
is neglected. 

From the viewpoint of the 4-dimensional model, 
we can easily write down the evolution of the 
energy density of $\Phi$ field as 
\begin{eqnarray}
\dot {\rho}^{(\Phi)} +4H {\rho}^{(\Phi)}+ 
  H\tau^{(\Phi)\mu}{}_{\mu} 
&=& -\Gamma_{\rm{esc}}\dot\Phi^2. 
\label{rhodot4d}
\end{eqnarray}
Then, with the aid of Eq.~(\ref{cons4dmodel}), we have 
\begin{eqnarray}
\dot {\rho}^{(D)} +4H {\rho}^{(D)}
&=& \Gamma_{\rm{esc}}\dot\Phi^2. 
\label{consdark}
\end{eqnarray}
Thus, we can naturally interpret 
that the $\Phi$ field decays into the dark radiation. 
Therefore, from the correspondence between the 4-dimensional and 
the 5-dimensional models, we are led to 
the interpretation that the dark radiation is generated 
as a result of this decay process in the 5-dimensional model. 

Usually, when we consider decay process, it leaves 
some decay products. However, the present 
decay process of the bulk inflaton
apparently does not have corresponding decay products,
since we do not consider the interaction with the other fields.
Then, one may think that 
once the bulk field decays in this manner, 
its effective energy density is simply lost 
without being transferred into another form. 
As was expected from the energy conservation law, 
this naive speculation is incorrect. 
In fact, the lost energy was shown to be    
transferred into the dark radiation component.  

Before closing this section, 
we mention that $\rho^{(\Phi)}$ in the 4-dimensional model 
does not correspond to $\rho^{(\phi)}=\rho^{(B)}$ 
in the 5-dimensional model. 
Hence, neither does $\rho^{(D)}$ to $\rho^{(E)}$ 
since $\rho^{(\Phi)}+\rho^{(D)}=\tilde\rho_{\rm eff}
   =\rho_{\rm eff}= \rho^{(\phi)}+\rho^{(E)}$. 
The fact that $\rho^{(\Phi)}\ne \rho^{(\phi)}$ 
will be easily seen by explicitly evaluating 
$\rho^{(B)}$ by using Eqs.(\ref{eq4}) and 
(\ref{EMtensor}) as
$
 \rho^{(\phi)} = (\ell/ 4)[
    \dot\phi^2+m^2\phi^2]
     = (1/ 4)\dot\Phi^2+(m_{\rm eff}^2/ 2)\Phi^2. 
$
This expression shows that 
${\rho}^{(\Phi)}={\rho}^{(\phi)}+\ell {\dot{\phi}}^{2}/4$. 

\subsection{Decay of the bulk scalar field to other fields}

In this section, we analyze the field equation (\ref{kgv})
without neglecting the dissipation terms on the background 
of the 5-dimensional anti-de Sitter space 
with a boundary de Sitter brane.
We make use of the fact that  
the late-time behavior is determined by the singularities
in the retarded Green function in the frequency domain,  
as was done in Ref.~\cite{newronbun}. 

The background metric can be written as 
\begin{equation}
ds^{2}=R(y)^{2}
\left(dy^{2}-d\tau^{2}+e^{2\tau}\,d\mbox{\boldmath
$x$}_{(3)}^{2}\right)
\,, 
\end{equation}
where $R(y)\equiv \ell\sinh^{-1}(|y|+y_0)$.  
The brane is located at $y=0$, 
and $y_0$ is specified by $\sinh y_0=H\ell$.   
The proper cosmological time on the brane $t$ is related to $\tau$ 
by $\tau \equiv Ht$. 
Setting $\phi=R^{-3/2}e^{-3\tau/2}\Psi(y,\tau)$, Eq.~(\ref{kgv}) 
reduces to 
\begin{equation}
  \left[-\partial_\tau^2+\partial_y^2-W(y)-U(y)\partial_\tau \right]\Psi 
  =0\, ,   
\label{fieldeq1}
\end{equation}
with 
\begin{eqnarray}
W(y) &=&{{15+4m^{2}\ell
^{2}}\over 4 \sinh^{2}(|y|+y_0)}\cr
 &&\hspace{1cm}-
{3\sqrt{1+H^{2}\ell^{2}}\over H\ell}\delta(y)-{3\over2}U(y),\\ 
U(y)&=& \Gamma_d \ell \, \delta\left(y \right)+\hat\Gamma_b(y) R(y)\,.
\end{eqnarray}
According to Ref.~\cite{newronbun}, the poles of the 
retarded Green function correspond to the solutions of 
Eq.~(\ref{fieldeq1}) that satisfy the outgoing boundary condition. 
The solutions can be found in the form of 
\begin{equation}
\Psi_{p} \propto e^{-ip \tau}  u_{p}.   
\label{eigen}
\end{equation}
Then, 
Eq.~(\ref{fieldeq1}) reduces to an eigenvalue equation
for $u_p$ having $p^2$ as the eigenvalue. 
When $\hat\Gamma_b=0$, 
the equation for $u_p$ can be solved 
with the outgoing boundary condition 
as 
\begin{equation}
  u_p (y) 
 = {1\over {\cal{N}}} P_{\nu-1/2}^{ip}[\coth(|y|+y_0)].
\label{upy}
\end{equation}
where 
$\nu:=\sqrt{m^2\ell^2+4}$. 
The normalization constant ${\cal{N}}$ is to be chosen 
so as to satisfy $\int_{-\infty}^{\infty} dy\, |u_p|^2 =1$.  
Here we have not yet imposed 
the boundary condition on the brane. 
The condition of the $Z_2$-symmetry 
\begin{equation}
\left[\left\{\partial_y+{3\over2}{\sqrt{1+H^{2}\ell^{2}}\over H\ell}
+\left(ip+{3\over2}\right){\Gamma_d \ell \over 2}\right\}
u_{p}(y)\right]_{y=0+}=0\,,
\label{pole}
\end{equation}
gives the equation that determines the eigenvalue, i.e., the 
location of the poles in the retarded Green function. 
We focus on the eigenmode with 
the eigenvalue $p$ that has the largest imaginary 
part in order to discuss the late-time behavior
of the bulk inflaton field.
Under the conditions (\ref{condition}), 
Eq. (\ref{pole}) for this eigenmode 
reduces to a quadratic equation as 
\begin{equation}
  p^2+\left(ip+{3\over 2}\right){\Gamma_d\over H}
          +{9\over 4}-{m^2\over 2 H^2}\approx 0. 
\label{standard}
\end{equation}

In order to illustrate the effect due to the dissipation in
the bulk, we consider two specific analytically solvable 
examples given by 
\begin{eqnarray}
 \hat\Gamma_b(y) R(y) &=& \Gamma_b H^{-1}\,,\quad [\mbox{case (I)}]\,, 
\label{gamma1}\\ \cr 
 \hat\Gamma_b(y) R(y) &=& 2\Gamma_b H R(y)^{2}\,,
      \quad [\mbox{case (II)}]\, 
\label{gamma2},
\end{eqnarray}
where $\Gamma_b$ is a constant. 
In case (I) the first term $p^2$ in Eq.~(\ref{standard}) is 
replaced with $p^2+(ip+3/2) \Gamma_b H^{-1}$, 
while in case (II) $m^2$ in the same equation 
is replaced with $m^2-2(ip+3/2) \Gamma_b H$. 
Then, in both cases we have the same equation 
to determine the eigenvalue of the largest imaginary part, 
and we find that a pair of such eigenvalues is given by 
\begin{eqnarray}
  p_{\pm} & \approx &  {1\over H}
\Biggl[-i{(\Gamma_d +\Gamma_b)\over 2}\cr
&&\quad \pm \sqrt{m^2_{\mathrm {eff}}-{(3H+\Gamma_d +\Gamma_b)^{2}
\over 4}}\Biggr]. 
\end{eqnarray}
These eigenvalues show that 
the time evolution of the bulk inflaton field evaluated on the brane 
at late times 
is identical to that of a 4-dimensional scalar field 
satisfying the equation obtained by replacing 
$\Gamma_{\rm esc}$ in Eq.~(\ref{infdynamics}) with 
$\Gamma_{\rm{tot}} = \Gamma_{\rm esc}+ \Gamma_d +\Gamma_b$.
Here we have added $\Gamma_{\rm esc}$ which comes from higher 
order corrections neglected in Eq.~(\ref{standard}). 
This result shows that all three decay channels 
damp the oscillation of the bulk inflaton field as is expected. 

So far we have studied just two specific examples.  
In more general cases,
the eigenfunction cannot be 
constructed analytically, but we can justify the extensive use of 
the above results by perturbatively
evaluating the shift of this pair of first eigenvalues.  
We consider that the unperturbed eigenfunction $u_p(y)$ 
is given by that for $U(y)=0$. 
Then applying the standard perturbation theory, we find 
that the shift of the eigenvalue $p^2$ 
due to the $U(y)$-term is given by 
\begin{eqnarray}
 \Delta p^2  & = &  -\left(ip+{3\over2}\right) \int \, dy \, 
   U(y) \vert u_p(y)\vert^2 \cr
 &\approx&  {-
 \displaystyle
   \left(ip+{3\over 2}\right)\left[\displaystyle\int_0^\infty dy \,
      \displaystyle{R(y) \hat\Gamma_b(y)\over (y+H\ell)^{3}}
     +{\Gamma_d\over 2(H\ell)^3}\right]
   \over \displaystyle \int_0^\infty {dy \over (y+H\ell)^{3}}}\, \,
\end{eqnarray}
where we have used the fact that 
$|u_p(y)|^{2}$ behaves as $R(y)^{3}\approx
[\ell/(|y|+H\ell)]^3$ near the 
brane. 
If we define $-(ip+3/2)\Gamma_b H^{-1}$ by 
the right hand side of the above equation, 
we recover the same results 
obtained above in the analytically solvable cases.
From the above equation, 
we also find that the dissipation is more efficient near the brane 
due to the effect of the warp factor.
This fact implies that 
the energy transferred to the other bulk fields 
is also possibly trapped around the brane as in the 
case of the $\phi$ field. 
Of course, however, 
if the bulk fields do not have degrees of freedom localized near 
the brane, the energy density near the brane will soon vanish.

Now let us consider the cosmic expansion law 
in the present case. 
We divide $T_{ab}$ into two parts as 
\begin{equation}
 T_{ab}= T_{ab}^{(\phi)}+T^{(*)}_{ab}\,. 
\end{equation}
Here $T^{(*)}_{ab}$ is the energy-momentum tensor of 
the other bulk fields, 
into which the bulk inflaton field $\phi$ may decay.
Then, the formulas for the energy density evolution 
of the matter fields on the brane (\ref{cons4d}) 
and that of the bulk fields (\ref{cons5d1}) 
reduce to 
\begin{eqnarray}
\dot\rho+4H\rho+H\tau^\mu_{~\mu}
&=& \left[\Gamma_d \dot \Phi^2  +2 {T}^{(*)}_{ab} u^a n^b
         \right]_{y=0}\,,
\label{rhodot}
\\
\dot \rho_{\rm{eff}} +4H \rho_{\rm{eff}} 
&=&\Biggl[
    -H\left(\dot\Phi^2-2m^2_{\mathrm{eff}}\Phi^2\right)\cr 
     &&\quad 
 -\Gamma_{d} \dot \Phi ^2  -2 H\ell \, {T}^{(*)}_{a b} n^a n^b
   \Biggr]_{y=0}\,,
\label{rhoeffdot}
\end{eqnarray}
where we used $\Phi$ defined by Eq.~(\ref{Phidef}). 
The cosmic expansion law depends on 
the nature of the fields composing $T_{ab}^{(\ast)}$. 
Below we discuss various cases one by one. 

\vspace*{2mm}
\centerline{\em 1) 
$T_{ab}^{(\ast)}\vert_{y=0} \approx 0$}
\vspace*{2mm}
Let us begin with the simplest situation 
that the bulk excitations caused by the 
decay of the $\phi$ field are not trapped around the brane. 
In this case $T_{ab}^{(\ast)}\vert_{y=0}$ will rapidly diminish, and hence 
Eqs.~(\ref{rhodot}) and (\ref{rhoeffdot}) become almost 
identical to those in the case without dissipation in the bulk. 
The main effect of $\Gamma_b$ arises in $\Gamma_{\rm{tot}}$,
and it can be absorbed by
the replacement of $\Gamma_{\rm esc}$ with $\Gamma_{\rm esc}+\Gamma_b$. 
Therefore we can immediately interpret 
that the $\Phi$-field decays into the dark radiation
with the decay rate $\Gamma_{\rm{esc}}+\Gamma_b$ 
and into the matter fields on the brane with the decay rate $\Gamma_d$.
If the matter fields on the brane stay radiationlike,  
the ratio of the dark radiation compared to the ordinary radiation 
is given by $(\Gamma_{\rm{esc}}+\Gamma_b)/\Gamma_d$. 
This ratio is constrained by the nucleosynthesis bound.
If $\rho$ decreases slower like dust matter or 
cosmological constant for a certain period, 
the constraint by the nucleosynthesis becomes weaker. 


\vspace*{2mm}
\centerline{\em 2) 
$T_{ab}^{(\ast)}\vert_{y=0} 
\approx \hspace{-9pt}
  /\,\, 0$ but ${T}^{(*)}_{ab} u^a n^b|_{y=0}= 0$}
\vspace*{2mm}
Next we consider the case that 
a part of resulting bulk excitation is trapped around the brane.
Then, $T_{ab}^{(\ast)}\vert_{y=0}$ 
will not diminish rapidly. Even in this case, 
if there is no direct interaction between the fields on the brane and 
the bulk fields other than $\phi$, 
we can set ${T}^{(*)}_{ab} u^a n^b|_{y=0}=0$, since
this term represents the energy transfer between them.
Therefore the effect of nonvanishing $T_{ab}^{(\ast)}\vert_{y=0}$ 
arises only in Eq.~(\ref{rhoeffdot}). 
Here the model qualitatively bifurcates into two cases,
depending on whether this bulk excitation 
is completely trapped or not. 

In the former case, the energy density of the trapped excitation
does not decay into anything. 
We cannot further analyze the evolution of this energy density 
without specifying the details of the model.  
In the latter case, the energy of the trapped excitation 
eventually escapes to the infinity, and as a result 
${T}^{(*)}_{a b} n^a n^b|_{y=0}$ vanishes in the end. 
As in the case of the $\phi$ field discussed in the preceding section, 
vanishing of ${T}^{(*)}_{a b} n^a n^b|_{y=0}$ will be 
interpreted as the transfer of 
the energy stored in the bulk modes into the 
dark radiation. 
As before, since the details of the model are not provided, we cannot analyze
the evolution of the energy density of this trapped excitation 
before it is transferred into the dark radiation. 
 
As a concrete example, let us consider the case that the decay 
product in the bulk is composed of the first pair 
of eigenmodes of another bulk scalar field.  
Then, since the effective energy density of this mode $\rho^{(*)}$ 
behaves like dust matter 
as in the case of the $\phi$ field, 
this field tends to dominate the energy density as 
a result of the cosmic expansion if the fields on the brane behave
as radiation. 
Therefore, models with bulk fields 
of this type will produce large amount of dark matter, which is possibly
transferred to dark radiation, 
and hence will be strongly constrained by the nucleosynthesis bound. 

\vspace*{2mm}
\centerline{\em 3) 
$T_{ab}^{(\ast)}\vert_{y=0} 
\approx \hspace{-9pt}
  /\,\, 0$ 
and ${T}^{(*)}_{ab} u^a n^b|_{y=0}
 \approx \hspace{-9pt}
  /\,\, 0$}
\vspace*{2mm}
Finally, we consider the case that there is 
bulk excitation trapped near the brane and it 
has direct interaction with the fields on the brane. 
In such models, most of the discussions given 
in the previous case apply in the same way.  
However, the history of the universe can be quite different,  
if this interaction is sufficiently strong. 
In this case, the trapped excitation 
can efficiently decay into the ordinary 
matter fields on the brane. 
Then, the fraction of dark component is reduced, and 
hence such models can avoid the 
conflict with the nucleosynthesis bound. 

\section{Summary}
We investigated braneworld models of Randall-Sundrum type 
with a massive bulk scalar field, 
which we call the bulk inflaton here, taking into account various
decay channels. 
We focused on how the cosmic expansion law realized on the brane is 
affected by the influence of the  
oscillating bulk inflaton and its decay products. 
The cosmic expansion law depends not only on the 
energy-momentum tensor 
of the fields localized on the brane but also on that of 
the bulk fields. 
Hence, in order to determine the cosmic expansion, 
we investigated the dynamics of the bulk inflaton 
including dissipation to the matter fields 
on the brane and in the bulk by using the Green 
function method developed 
in Refs.\cite{Himemoto:2001hu,newronbun,Dubovsky:2000am}. 

It was argued that the effect of the bulk inflaton on the 
cosmic expansion can be mimicked by an effective 4-dimensional inflaton 
as long as $H^{2}\ell^{2} \ll 1$ and 
$|m^{2}|\ell^{2} \ll 1$ in Ref.~\cite{Himemoto:2001hu}. 
Examining the late-time behavior of solutions of the bulk 
scalar field, 
we have shown that this claim is valid 
even if the bulk inflaton has various decay channels. 
The decay of the bulk inflaton occurs not only due to the 
dissipation to other fields on the brane or in the bulk 
but also due to the escape of energy to the direction 
away from the brane. 
From the analysis of the cosmic expansion law, 
we have shown that the decay product of this latter intrinsic 
dissipation should be interpreted as the dark radiation 
from the viewpoint of 4-dimensional effective theory. 
Further, we stress that this dark radiation is not 
identical to the contribution from the 
projected 5-dimensional Weyl tensor. 


As for the dissipation due to interaction with matter fields 
on the brane, it has been already shown in Ref.~\cite{newronbun} that 
the energy lost by this decay process 
can be interpreted as simply transferred to the matter fields on the brane, 
which is analogous to the usual decay process of 
the 4-dimensional inflaton field. 
The consequence of the decay due to interaction with other bulk 
fields has vast varieties. If the excitation generated by 
this decay process is not trapped near the brane at all, the effect is 
equivalent to the case of intrinsic dissipation. 
However, we also found that the dissipation to the bulk 
is more efficient near the brane. 
Hence, it is likely that a part of the lost energy 
is transferred to a mode localized near the brane. 
In this case, the decay 
product rather behaves as a 4-dimensional 
field as in the case of the bulk inflaton. 
The effective equation of state of such excitation measured 
through the cosmic expansion law depends on 
the details of the model setup, 
but at least we know that the effective equation of state is 
dustlike if the decay product consists of 
another massive scalar field. 
If the equation of state is dustlike, the fraction of the 
energy density of this excitation continues to increase 
during the radiation dominant phase. 
In the case that this excitation does not decay into the 
fields on the brane, it behaves as ``dark matter''. 
If this excitation is not completely trapped 
in the neighborhood of the brane, it finally decays into 
the dark radiation in the same way as we have seen for the bulk inflaton. 
This further decay transfers the energy from this ``dark matter'' 
to the dark radiation, but the 
dark component anyway in total dominates the energy density 
unless the decay rate of this excitation is sufficiently high or 
its initial amplitude is extremely small. 
Hence, such kind of models will be severely 
constrained by the requirement of the big-bang nucleosynthesis. 

If the excitation in the bulk generated by the bulk inflaton decay  
has interaction with the fields on the brane, 
the energy can be transferred to the ordinary radiation fluid. 
If such decay process is efficient enough compared to the other 
decay channels, the final fraction of the dark component will not 
be significantly large. Such kind of models will be viable. 

In conclusion, we found that the bulk inflaton model 
can reproduce the standard cosmology, but 
the branching ratio 
of the bulk inflaton decay and the nature of the decay products 
are constrained. 
The bulk inflaton model might leave observable signatures of 
the braneworld, and it may also provide a key to 
understanding the origin of dark matter. 
%

\section*{Acknowledgments} 
To complete this work, the discussion during and 
after the YITP workshops YITP-W-01-15 and YITP-W-02-19  
was useful. This work was supported in part 
by the Monbukagakusho Grant-in-Aid 
Nos.~14740165 and 14047212.

\end{multicols}
\end{document}